\documentclass{aa}
\usepackage{graphicx}
\usepackage[varg]{txfonts}
\usepackage{enumitem}
\usepackage{amsmath,amssymb}
\usepackage{booktabs}
\usepackage{xspace}
\usepackage{textcomp}
\usepackage{xcolor}	
\usepackage[euler]{textgreek}
\usepackage{dirtytalk}
\usepackage[colorlinks=true, linktocpage, linkcolor={blue!60!black}, citecolor={blue!60!black}, urlcolor={blue!60!black}]{hyperref}

\usepackage{caption}
\usepackage{subcaption}

\newcommand{\oiii}{[\ion{O}{III}]\,88\,\textmu m~}
\newcommand{\cii}{[\ion{C}{II}]\,158\,\textmu m~}

\title{Quantifying the detection likelihood of faint peaks in interferometric data through jackknifing} 

\subtitle{Test application on finding $z>10$ galaxy candidates}

\titlerunning{Detection likelihood of faint peaks in interferometric data}

\author{
Joshiwa van Marrewijk\inst{1,2}, Melanie Kaasinen\inst{1}, Gerg\"o Popping\inst{1}, Luca Di Mascolo\inst{3,4}, Tony Mroczkowski\inst{1}, Leindert Boogaard\inst{2,5}, Francesco Valentino\inst{1}, Tom Bakx\inst{6}, Ilsang Yoon\inst{7}}

\authorrunning{J. van Marrewijk, M. Kaasinen, et al.}    

\institute{
          European Southern Observatory, Karl-Schwarzschild-Str. 2, D-85748, Garching, Germany
    \and
          Leiden Observatory, Leiden University, P.O. Box 9513, 2300 RA Leiden, The Netherlands \\
        \email{marrewijk@strw.leidenuniv.nl}
    \and
          Laboratoire Lagrange, Université Côte d'Azur, Observatoire de la Côte d'Azur, CNRS, Blvd de l'Observatoire, CS 34229, 06304 Nice cedex 4, France
    \and
        Kapteyn Astronomical Institute, University of Groningen, Landleven 12, 9747 AD Groningen, The Netherlands
    \and
        Max Planck Institute for Astronomy, K\"onigstuhl 17, 69117 Heidelberg, Germany
    \and 
        Department of Space, Earth, \& Environment, Chalmers University of Technology, Chalmersplatsen 4 412 96 Gothenburg, Sweden
    \and 
        National Radio Astronomy Observatory, 520 Edgemont Road, Charlottesville, VA 22903, USA
}

\date{Received 20/08/2024; accepted ...}

\abstract
{False-positive emission-line detections bias our understanding of astronomical sources; for example, falsely identifying $z\sim3-4$ passive galaxies as $z>10$ galaxies leads to incorrect number counts and flawed tests of cosmology.}
{In this work, we provide a novel but simple tool to better quantify the detection of faint lines in interferometric data sets and properly characterize the underlying noise distribution. We demonstrate the method on three sets of archival observations of $z>10$ galaxy candidates, taken with the Atacama Large Millimeter/Submillimeter Array (ALMA).}
{By jackknifing the visibilities using our tool, \texttt{jackknify}, we create observation-specific noise realizations of the interferometric measurement set. We apply a line-finding algorithm to both the noise cubes and the real data and determine the likelihood that any given positive peak is a real signal by taking the ratio of the two sampled probability distributions.
}
{We show that the previously reported, tentative emission-line detections of these $z>10$ galaxy candidates are consistent with noise. 
We further expand upon the technique and demonstrate how to properly incorporate prior information on the redshift of the candidate from auxiliary data, such as from JWST. 
}
{Our work highlights the need to achieve a significance of $\gtrsim 5\sigma$ to confirm an emission line when searching in broad 30 GHz bandwidths. Using our publicly available method (\url{https://joshiwavm.github.io/jackknify/}) enables the quantification of false detection likelihoods, which are crucial for accurately interpreting line detections.}


\begin{document}

\maketitle
%

\section{Introduction}

    \begin{table*}[t]
        \begin{center}
        {
        \begin{tabular}{@{}lccc@{}}
            \multicolumn{4}{c}{\textbf{Archival ALMA observations of $z>10$ galaxy candidates}} \\ 
            \toprule
             \textbf{ID}                    & GLASS-z12                                            & GLASS-z10                            & S5-z17-1 \\
             \midrule
             RA [deg]                       & 3.4990                                               & 3.5119                               & 339.0160 \\
             DEC [deg]                      & -30.3248                                             & -30.3719                             & 33.9046 \\
             ALMA source name               & GLASS-z13 [GLASS-z12]                                & GLz11                                & S5-z17-1 \\
             PI                             & T. Bakx [J. Zavala]                                  & I. Yoon                              & S. Fujimoto \\
            ALMA Project Code               & 2021.A.00020.S [2023.A.00017.S]                      & 2021.A.00023.S                       & 2021.A.00031.S \\
            ALMA Band                       & 6 [6]                                                & 7                                    & 7 \\
            Beam [\arcsec]                  & $0\farcs32\times0\farcs29, 46^\circ$  [$0\farcs45\times0\farcs40, 73^\circ$]          & $0\farcs77\times0\farcs59, 81^\circ$ & $0\farcs78\times0\farcs45, 3^\circ$ \\
            Channel width [MHz]             & 31 [31]                                              & 31                                   & 31   \\
            On source time [h]              & 8.1 [1.6]                                            & 6.9                                  & 0.5  \\
            \midrule
            \textbf{Reported tentative detections} & \\
            \midrule
            Ref. for initial identification & \citet{Castellano2022},\citet{Naidu2022}          & \citet{Naidu2022}                    & \citet{Harikane2023} \\
            Reference for ALMA data         & \citet{Bakx2023}, \citet{Zavala2024b,Zavala2024}$^\dagger$   & \citet{Yoon2023}                     & \citet{Fujimoto2023} \\
            $z_\mathrm{phot}$               & $12.2\pm 0.2$                                     &  $10.4\pm 0.5$                       & $16.7^{+1.9}_{-0.3}$ \\
            $\Delta z$ covered              & $11.9 - 13.5$                                     & $10.1 - 11.1$                        & $16.1 - 17.3$ \\
            S/N (\oiii)$^{*}$               &  5.8, 3                                           & 4.4                                  & 5.1 \\
            Offset                          &  $0\farcs5$,  [$0\farcs0$]                        & $0\farcs17$                          & -- \\
            $z_\mathrm{spec}^{ALMA}$        &  $12.117 \pm 0.001$,  [$12.3327 \pm 0.0035$]      & 10.38                                & 16.01 \\
            $z_\mathrm{spec}^{JWST}$        &  $12.33 \pm 0.02$                                 & --                                   & -- \\
            Freq [GHz]                      & 258.7           ,  [$254.487 \pm 0.019$]          & 298.25                               & 338.726 \\
            $\Delta \nu$ [km s$^{-1}$]      & $400\pm 70$, [$186\pm 58$]                        & 225                                  & 118 \\
            \bottomrule 
        \end{tabular}
        }
        \caption{Candidate coordinates and reported detection significance of the previously reported line searches. $^\dagger$The reported details after precise redshift information became available from JWST observations and deeper ALMA Cycle 11 observations and shown within brackets. $^*$ In the case of S5-z17-1, we quote the redshift of the plausible [\ion{O}{III}]\,52\,\textmu m line. The beams are given in terms of major and minor beam axes and position angle.}
        \label{tab:observations}
        \end{center}
    \end{table*}

    Assessing the reliability of faint astronomical signals is essential for all areas of astrophysics and cosmology. To understand the entirety of a population, the astronomical community is always attempting to characterize sources at the limits of what can be observed with the current generation of facilities. Interferometers such as the Atacama Large Millimeter/Submillimeter Array (ALMA) and the Very Large Array (VLA) have enabled the efficient study of faint, distant sources in the mm/cm-wave regime. However, images generated from interferometric data are highly complex; their underlying noise distribution is challenging to quantify, leading to potential biases at low $S/N$ and difficulties in interpreting the data. This can have profound consequences, for example, when determining the existence, redshift, and physical properties of faint galaxies.
    
    Correctly determining the likelihood of faint peaks in interferometric data has become relevant in the context of studies of the interstellar medium in the first galaxies ($z>10$). Over the past two years, six ALMA Director's Discretionary Time (DDT) programs\footnote{Additionally, one 80+ hour NOEMA observation on GN-z11 \citep[which resulted in an upper limit;][]{Fudamoto2024}} were approved and executed to follow up $z>10$ galaxy candidates. These galaxies were initially identified from optical and near-infrared photometry, mostly including data from the JWST. However, modeling the spectral line energy distribution of these photometric data resulted in a wide redshift probability distribution, including several potential solutions. Capitalizing on the broad spectral coverage of ALMA, these DDT programs aimed to confirm the redshifts of the galaxy candidates by targeting the \oiii line. But, obtaining robust line detections with an accurate $S/N$ estimate has proven challenging.
        
    So far, ALMA DDT observations of $z>10$ candidates have resulted in a handful of potential upper limits on the targeted line emission \citep{Popping2023, Kaasinen2023} and marginal, low $S/N$ ($3-4\sigma$) detections of \oiii (see, \citealt{Harikane2022, Yoon2023, Bakx2023, Fujimoto2023}), with the exception of one source, JADES-GS-z14, for which a $6\sigma$ detection of the \oiii line was recently reported \citep{Carniani2024b, Carniani2024, Schouws2024}. Some of these marginal detections have since been proven not to correspond to actual emission lines, with follow-up JWST/MRS and JWST/NIRSpec spectra revealing at least two of these galaxies to be at a different redshift to that implied by the initial false positive \oiii\ detection \citep{Zavala2024, Harikane2024b}. This highlights a broader issue: false positive identifications of emission lines lead to biases in the derived physical properties. In the case of incorrect redshift solutions, this has profound implications for the number counts of the earliest galaxies, thereby impacting our understanding of both cosmology and galaxy evolution.

    In this work, we present a straightforward and effective technique to quantify the detection level and underlying noise distribution in interferometric data sets. We build upon the approach used in \citet{Kaasinen2023}; by differencing the visibilities, we generate various noise realizations of the observation-specific ALMA measurement set. Then, by applying line-finding algorithms on both the real and source-free image cubes, we sample the likelihood of a marginal detection being real without needing to assume an underlying noise distribution. We make our technique readily available to the community in the form of the public tool \texttt{jackknify}, along with clear, step-by-step tutorials.\footnote{see \url{https://joshiwavm.github.io/jackknify/}} Although we implement and optimize the tool for ALMA data, \texttt{jackknify} is compatible with any type of interferometric observations that use the \textit{Common Astronomy Software Applications for Radio Astronomy} \citep[CASA,][]{CASA2022}. For example, this jackknifing technique may be useful for characterizing the faint HI emission from $z>1$ galaxies using Squared Kilometer Array \citep[SKA;][]{Dewdney2009} and its precursors \citep[see, e.g., ][]{Baker2024},\footnote{However, we do note that our tools are validated only on data where the $w$-term was neglected.} or quantifying the faint molecular line emission from quiescent galaxies with ALMA. In this work, we apply \texttt{jackknify} to three archival DDT observations targeting $z>10$ galaxy candidates. We will test two scenarios: one in which no prior knowledge regarding the source's redshift is available and one in which we can use a prior from JWST on the frequency location of the line. 
    
    The paper is outlined as follows. Section~\ref{sec:observations} describes which interferometric data sets we use, Section~\ref{sec:jack_knife} provides the overview of how \texttt{jackknify} is built, Section~\ref{sec:results_sims} describes the performance of the tool via simulated observations, Section~\ref{sec:results} contains the results of applying \texttt{jackknify} to the ALMA observations of $z>10$ galaxy candidates, and Section \ref{sec:summary} provides an overview and summary of this work. For all calculations, the assumed cosmology is based on \citet{Planck2014_XIII}, a spatially flat, $\Lambda $ cold dark matter ($\Lambda$CDM) model with $H_0=67.7$\, kms$^{-1}$~Mpc$^{-1}$ and $\Omega_{\rm m} = 0.307$.

\section{The archival ALMA observations} \label{sec:observations}

        We test and validate our jackknifing tool, \texttt{jackknify}, on three sets of ALMA DDT observations targeting $z>10$ galaxy candidates pre-identified from photometry \citep[as presented in][]{Castellano2022, Naidu2022, Harikane2023}. All of these DDT observations \citep{Bakx2023, Yoon2023, Fujimoto2023} aimed to detect the \oiii emission line over a wide redshift probability range. To maximize the redshift coverage, each spectral setup consisted of four adjacent tunings, covering a large, $\approx 30$~GHz bandwidth. A comprehensive overview of the literature results and observational specifics is presented in Table~\ref{tab:observations}.

        Calibrated measurement sets were provided by the European ALMA Regional Centre network \citep{Hatziminaoglou2015} through the calMS service \citep[CalMS;][]{Petry2020} using CASA v5.4.0. To reduce the data size, we performed two additional operations on the calibrated data, applying a time averaging of 30 seconds and spectral averaging of the visibilities in $\approx 30$ MHz bins. The frequency binning was conducted in a manner that avoided interpolation over the frequency axis. 
        We used \texttt{uvcontsub} to manually continuum subtract a first-order polynomial from the measurement sets to remove any potential continuum emitters. We omitted the frequency channels where tentative features might be present in the data based on the previously reported detections. 
        Finally, we imaged all measurement sets using the \textsc{tclean} task with natural weighting. 

\section{Methodology}\label{sec:jack_knife}

    \subsection{Jackknifing}

        \begin{figure}[t]
            \centering
            \includegraphics[width=\hsize]{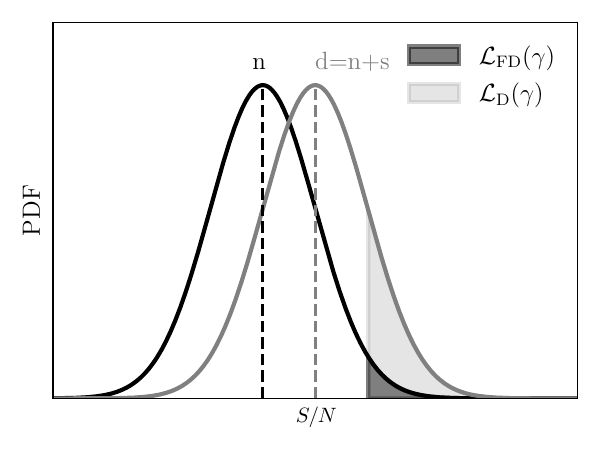}
            \caption{Schematic of the detection inference. Given the data, $d$, which is the linear combination of the noise distribution, $n$, plus the signal, $s$, we can define the likelihood of detection $\mathcal{L_{\rm D}}(\gamma)$ by setting an arbitrary $S/N$ threshold $\gamma$ and integrating the probability density function of the $S/N$ for $d$ from $\gamma$ to $\infty$, as indicated with the shaded area. By nullifying $s$ through jackknifing of $d$, we recover the ideal thermal noise, $n$, and thus compute the likelihood of a false positive detection $\mathcal{L_{\rm FD}}(\gamma)$ as we do for computing $\mathcal{L_{\rm D}}(\gamma)$. The ratio of the two likelihoods provides the significance of detection. This figure is inspired by Fig.~1 of \citet{Vio2016}.}
            \label{fig:schematic}
        \end{figure}

        In the narrow-field approximation, interferometers measure the 2D Fourier transform of the intensity distribution of an astronomical source over a sparse collection of Fourier modes. Such measurement modes, called visibilities, have units of wavenumbers and are expressed in terms of \textit{uv}-distances, where $u$ and $v$ are the orthogonal vector bases of the Fourier space. Assuming proper calibration\footnote{In the case of incorrect calibration, such as when one of the 44 online ALMA antennas has an erroneous gain solution, the noise in the visibility plane becomes multiplicative rather than additive to the signal. The effects of jackknifing in such a situation are not thoroughly tested. However, the assumption of well-calibrated data without antenna calibration errors is reasonable, as it is common practice to check the calibration accuracy before imaging to ensure there are no spurious signals that could result in imaging artifacts.}, noise in the Fourier domain, unlike in the image plane, is additive and Gaussian. This occurs because the individual pulses from electrons are so numerous that they blend into an indistinguishable, continuous Gaussian random process with a mean of zero \citep{Thompson2017}.

        In this work, we use jackknifing to characterize the noise properties of the data. Jackknifing is a common technique used to retrieve the noise properties from bolometric measurements taken with mm-wave single-dish facilities. It is most commonly employed on Cosmic Microwave Background (CMB)/Single dish experiment observations \citep[e.g.,][]{Weiss2009, Romero2018, Naess2020}; however, it has also been used extensively to characterize radio/mm-wave interferometric data \citep[e.g.,][]{Padin2001, Readhead2004, Sharp2010, Keating2015}. Since interferometric data has uncorrelated, additive, and Gaussian noise with a mean of zero, we can retrieve the uncorrelated noise distribution, $n$, from the data, $d$, by jackknifing -- by randomly multiplying half of the real and imaginary amplitudes of the visibilities by -1 and then rebinning, thereby averaging out the signal, $s$ \citep[see, for instance,][for an effective implementation]{Kitayama2020, GonzalezL2020, Kaasinen2023, DiMascolo2023}. 

        When visibilities are imaged, the $uv$-data is automatically rebinned and averaged since all imaging and deconvolution tools, including \texttt{tclean}, grid the $uv$-coordinates before Fourier transforming. Any artifacts caused by the imaging, such as spatially correlated noise, are captured by performing the jackknifing in the visibility plane and performing the inference in the image plane. Jackknifing in the visibility plane is thus essential, rather than splitting in the image plane, to avoid removing any correlated noise induced by the Fourier transform present in the data. 
        
        To generate various noise realizations, we change the seeding of the randomization process. This allows us to sample the noise distribution, $n$, until convergence. 
        Once $n$ is adequately sampled, we use the distribution to infer the likelihood of false detection. Figure~\ref{fig:schematic} shows schematically how this type of detection inference works. Given that the data, $d$, is a linear combination of the noise distribution, $n$, plus the signal, $s$, we can define the likelihood of detection $\mathcal{L_{\rm D}}(\gamma)$ by integrating from an arbitrary $S/N$ threshold, $\gamma$, onward, as indicated with the shaded area in Figure~\ref{fig:schematic}. By nullifying $s$ through jackknifing of $d$, we recover $n$ in the visibility plane and thus are able to compute the likelihood of a false positive detection $\mathcal{L_{\rm FD}}(\gamma)$. The ratio of the two likelihoods provides the significance of detection. 

        Resampling the noise distributions requires that the cubes are re-imaged at every iteration. We ensure that \texttt{jackknify} interfaces with \texttt{CASA} and \texttt{tclean} to retain similar metadata for the spectral cubes, such that it is compatible with most line-finding algorithms. Because of this, \texttt{jackknify} can become computationally intensive, particularly for large data sets. As we will show in this paper, the analysis of ALMA spectral scans is computationally feasible. Large mosaics, however, might be more problematic. Therefore, we have created another implementation named \textit{jaxknify}, which utilizes the \texttt{jax} implementation of the \texttt{iFFT} as a substitute for \texttt{tclean}. The output of the Fourier Transform is, however, not compatible with existing line-find methods. Hence, for the remainder of this work, we used the \texttt{CASA}-compatible version of \texttt{jackknify}.\footnote{\texttt{jaxknify} was only used to compute the many realizations which are visualized in Figure~\ref{fig:test_simobs} as described in Section~\ref{sec:results_sims}.} 

    \subsection{Line finding}

        \subsubsection{Traditional empirical methods}\label{sec:trad}

            To date, most approaches to finding emission-line sources in (sub-)mm interferometric data have been based on determining the significance of an emission line (a positive peak) via the line ``fidelity'' or ``reliability'' \citep[e.g.,][]{Walter2016,Pavesi_2018, GL2019,Westmeier_2021}. 
            The fidelity (or reliability) function is defined as,
            
            \begin{equation}\label{eq:fidelity}
                \mathrm{fidelity (\gamma)} = 1 - \dfrac{N_\mathrm{neg} (\mathrm{\gamma})}{N_\mathrm{pos} (\mathrm{\gamma})}~,
            \end{equation}
            
            \noindent where $N_\mathrm{neg} (\mathrm{\gamma})$ and $N_\mathrm{pos} (\mathrm{\gamma})$ are the number of negative and positive peaks, respectively, above a given detection threshold $\gamma$. The value of $\gamma$ is usually a function of the $S/N$ but can be arbitrarily chosen. 
            
            The fidelity function empirically estimates the likelihood ratio of the false detection, $\mathcal{L_{\rm FD}}(\gamma)$, over the detection, $\mathcal{L_{\rm D}}(\gamma)$ by taking the integral over the Probability Density Functions (PDFs) of their respective probability distributions, $\mathcal{P_{\rm FD}}(x)$ and $\mathcal{P_{\rm D}}(x)$. $\mathcal{P_{\rm FD}}(x)$ and $\mathcal{P_{\rm D}}(x)$ are sampled by searching for negative and positive peaks, respectively. Therefore, the following relation of the likelihood ratio between detection and false detection holds:
            \begin{equation}\label{eq:ratio}
                \Lambda(\gamma) = \dfrac{N_\mathrm{pos} (\mathrm{\gamma})}{N_\mathrm{neg}(\mathrm{\gamma})} = \dfrac{\mathcal{L_{\rm D}}(\gamma)}{\mathcal{L_{\rm FD}}(\gamma)} = \dfrac{ \int_{\gamma}^{\infty} \mathcal{P}_{\rm D}(x | \Delta v)~\mathrm{d}x}{ \int_{\gamma}^{\infty} \mathcal{P}_{\rm FD}(x | \Delta v)~\mathrm{d}x}~.
            \end{equation}

            \noindent This relationship is a function of the kernel width used to determine the linewidth, $\Delta v$. For example, to characterize the line emission in the ALMA Spectroscopic Survey of the Hubble Ultra Deep Field (ASPECS) \cite{GL2019} chose the detection threshold $\gamma$ such that $\mathrm{fidelity (\gamma)}$ = 0.9. This resulted in $\gamma \simeq 6.4\sigma$ for the ASPECS sample, assuming that Eq.~\eqref{eq:fidelity} is well described by an error function \citep[see also,][]{Decarli2020}. 
    
            The other assumption made while employing the fidelity function is that the flux distribution of negative peak values is a proper estimator for $\mathcal{P}_{\rm FD}(x)$. However, this does not hold when absorption with respect to the CMB or a bright background source leads to negative line fluxes, contaminating the noise statistics. Additionally, since the method is applied in the image domain, any bright line or continuum flux will also create a negative signal through the dirty beam patterns. Bright sources can be ``cleaned'' (if they fall within the imaged part of the sky), but the fainter ones will create correlated spatial noise \citep{Tsukui2022}. Furthermore, ALMA has a limited data volume along the spectral axis; hence, $\mathcal{P}_{\rm FD}(x)$ is not fully sampled over the $\delta\nu$ space, making these empirical approaches inaccurate.

        \subsubsection{Line detection inference through jackknifing}\label{sec:jackfinding}

            Here, we describe how we perform the line detection throughout this work by utilizing the jackknifed data sets. To determine the likelihood of detection, we first quantify the likelihood of false detection, for which we need the underlying PDF. The distribution of false positive detections is set by the distribution of peak values in a data cube that only contains noise (Section~\ref{sec:trad}). The distribution of positive or negative peaks is a subset of the pixel-value distribution in the data cube \citep[][]{Vio2016}. The number, location, and amplitude of peak values depend on the realization of the total pixel-value distribution, which for interferometric data in the image plane, can be approximated as a smoothed random Gaussian field. By jackknifing the visibilities, we effectively shuffle the noise distribution, altering the noise realization and creating another subsample of the pixel-value distribution, namely another peak distribution. This allows us to increase the sample statistics of $\mathcal{P}_{\rm FD}(x)$ while using the same data set.

\begin{figure}[t!]
                \centering
                \includegraphics[width=\hsize]{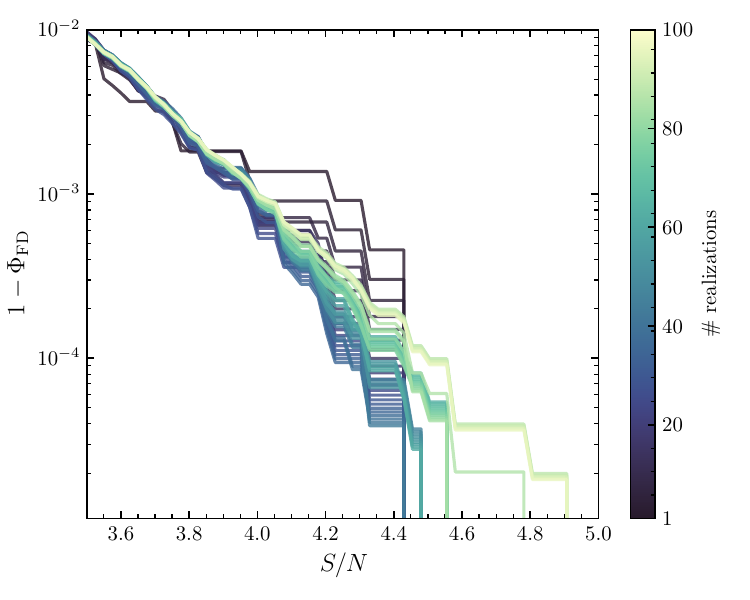}
                \caption{How an increasing number of jackknife realizations leads to a smoother $\Phi(x)$ that better samples the high-$S/N$ of the probability of false detection, $\mathcal{P}_{\rm FD}(x)$. The probability functions are shown as one minus the mean cumulative distributions $\Phi(x)$ of $\mathcal{P}_{\rm FD}(x)$ as a function of the peak $S/N$ for various amounts of jackknife realizations.}
                \label{fig:convergence}
            \end{figure}

            Figure~\ref{fig:convergence} shows that for an increasing number of realizations, we better probe the tail of the distribution at high $S/N$. The cumulative distribution function (CDF) $\Phi_{\rm FD}$ of $\mathcal{P}_{\rm FD}(x)$ is smooth till $S/N\approx4.5$ and reaches a $S/N = 5$ within a hundred realizations. Specifically, each CDF incrementally includes data from additional jackknife realizations. For example, the CDF for $35\times$ jackknife realizations contains all the data from the $34\times$ realization set. Furthermore, the jackknifed data only includes peaks within a $0\farcs6$ aperture of the center of the data cube (chosen so that this test is consistent with the discussion in section~\ref{sec:results}). Figure~\ref{fig:convergence} extends to a hundred realizations in total. Since there is a trade-off between sampling more realizations and computation time and from $\approx$ 50 realizations onward we recover the high-$S/N$ tail of the distribution; we decided to use 50 jackknife realizations throughout this work. 
            
\begin{figure}[t!]
                \centering
                \includegraphics[width = 0.49\textwidth]{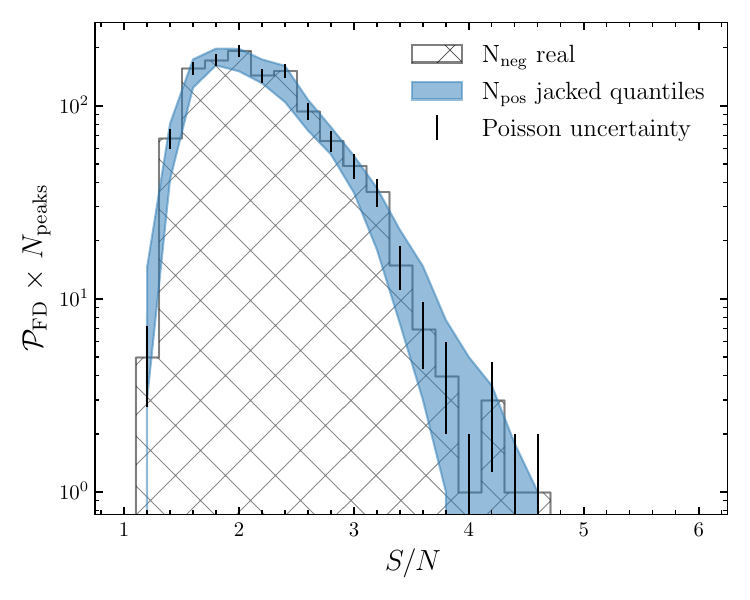}
                \caption{Sampled noise distributions, $\mathcal{P}_{\rm FD}(x)$, as a function of the peak $S/N$. The gray hatched histogram shows the results from sampling the number of negative peak values of the original data set (shown with corresponding Poisson uncertainty). The blue-filled region represents the 95\% confidence interval of the positive peak values from the jackknife observations.}
                \label{fig:pdf_noise}
            \end{figure}

            The difference between the sampled PDF of the false positives, $\mathcal{P_{\rm FD}}(x)$, coming from the negative peak distribution (e.g., one noise realization) in the real data versus that from 50 jackknife realizations is shown in Figure~\ref{fig:pdf_noise}. For the jackknifed realizations, we show the inner 95\% confidence interval per bin. Again, for both the real and jackknifed data, we only consider peaks within a $0\farcs6$ aperture of the center of the data cube. Since we do not expect a negative signal in this data set, the two distributions should be equivalent. However, our method provides better sampling (as also shown in Figure~\ref{fig:convergence}) and thus better statistics, particularly at higher $S/N$. Furthermore, the scatter in the jackknife realizations within the 95\% confidence interval is consistent with the Poisson uncertainty in the real data for each bin, except for a potential outlier at $S/N\approx 4$. This indicates that the various realizations are consistent with each other.

            As shown in Figures~\ref{fig:convergence}~\&~\ref{fig:pdf_noise}, jackknifing allows for a more complete sampling of the noise distribution without the need for complex models or computationally intensive four-dimensional covariance estimates \citep[see,][for a discussion on the autocorrelation function implementations used for linefinding]{Vio2016, Tsukui2022}. By leveraging \texttt{jackknify}, we can obtain a more reliable and efficient measure of the probability of false detection, $\mathcal{P_{\rm FD}}(x)$, and take the likelihood ratio, $\Lambda(\gamma)$ with $\mathcal{L_{\rm D}}(\gamma)$ according to Eq.~\eqref{eq:ratio} to estimate a feature being real or not. Then,             
            \begin{itemize}
                \item if \(\Lambda(\gamma) > 1\), the observed data is more likely to fall under the hypothesis that a signal is present, whereas
                \item if \(\Lambda(\gamma) \leq 1\), the observed data falls under the null hypothesis and is thus better described as being drawn from the underlying noise distribution.
            \end{itemize}
            \noindent Finally, an excess in the ratio is considered significant if the likelihood ratio exceeds a critical value \(k\), such that:
            \begin{itemize}
                \item if \(\Lambda(\gamma) \geq k\), one can reject the null hypothesis that no signal is present in the data, whereas
                \item if \(\Lambda(\gamma) < k\), one cannot reject the null hypothesis.
            \end{itemize}

\section{Application to simulated ALMA data}\label{sec:results_sims}
                        
    \begin{figure*}[t!]
        \centering
        \includegraphics[width=0.95\textwidth]{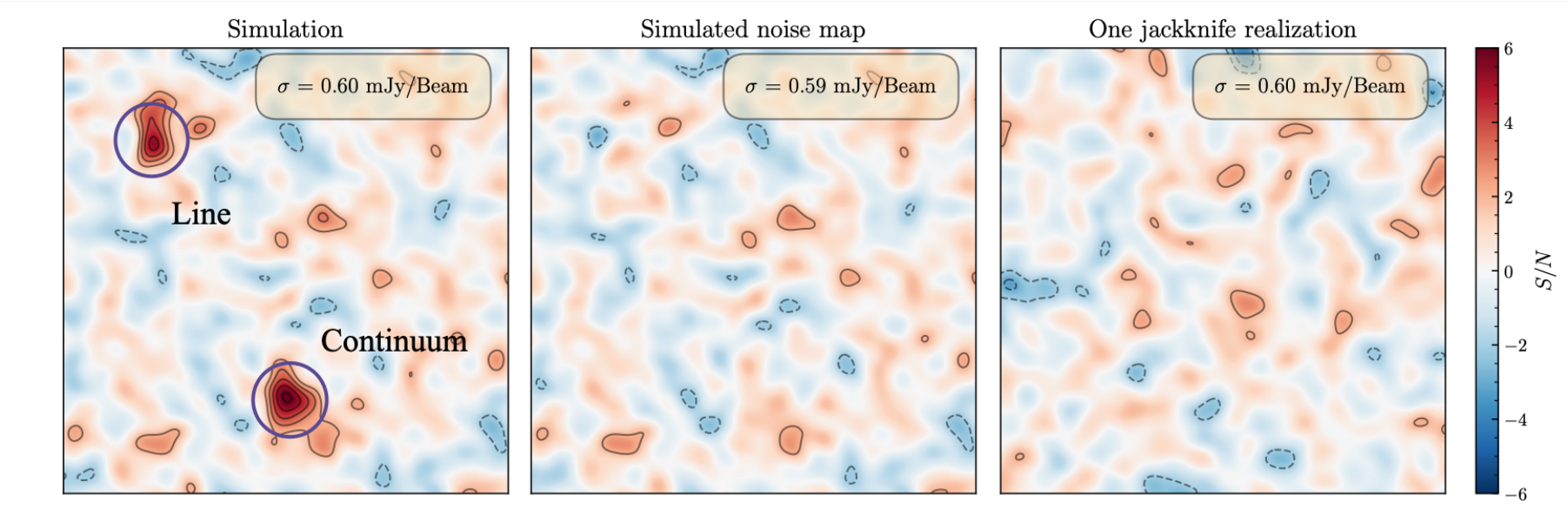}
        \caption{Moment-0 maps of simulated ALMA data, generated using \texttt{CASA simobserve}. Left: Output of the simulated observation, which contains both a continuum and a line-emitting source with $S/N=5$. Middle: The corresponding noise map, referred to as the ``cleaned'' map, obtained by subtracting the noiseless output (i.e., filtered sky model) of \texttt{simobserve} from the noisy one. Right: The moment-0 map from a single jackknife realization. All contours are drawn at $[\pm2, \pm3,\pm4,\pm5,\pm6]\,\sigma$.}
        \label{fig:sims}
    \end{figure*}
            
    To validate our method, we simulated mock ALMA data using \texttt{CASA}'s \texttt{simobserve} task. We simulated a one-hour-long execution block centered at an RA and Dec of zero degrees. The simulated data contained six channels centered at a frequency of $\nu = 279$~GHz with a channel width of 31~MHz. We created the simulated $uv$-coverage using configuration 4 from cycle 7, leading to a resolution of $0\farcs6$ when \texttt{cleaning} using natural weighting. The setup of the simulations was designed to be similar to the observational setup of GLASS-z12 (Table~\ref{tab:observations}). Using this setup, we simulated several datasets containing continuum and line emission from two separate sources with various integrated $S/N$ values. We incremented the flux of the sources, maintaining the same $uv$-coverage for every observation. The line flux was simulated using a 3D Gaussian model. One dimension corresponds to the spectral flux distribution; its flux is constrained as described above -- so that the integral over the Gaussian equals the respective $S/N$ -- and has an FWHM equal to $\sim73$~MHz, corresponding to a line with an FWHM of 80 km s$^{-1}$. The second and third dimensions of the Gaussian correspond to the spatial dimensions and are used to generate a blob in each channel map, with its amplitude set by the first Gaussian and an FWHM of 2.355 pixels (1 pixel = $0\farcs01$). Hence, the observation is considered unresolved. The same angular size on the sky is used for the continuum source.

    Using the simulated data, we checked how well \texttt{jackknify} removes the signal. Figure~\ref{fig:sims} shows the moment-0 maps of the $S/N = 5$ observation (\textit{Left}), the corresponding noise map (\textit{Middle}), and the map resulting from jackknifing the visibilities (\textit{Right}). 
    We created the noise-only map by subtracting the noise-free dirty map (which is an output of \texttt{simobserve}) from the noisy output. Thus, the noise-only map is representative of the noise in a ``perfectly \texttt{cleaned}'' map. The simulated noise map and jackknifed realization exhibit the same noise structure, validating our jackknifing approach.

\begin{figure}[t!]
        \centering
        \includegraphics[width=\hsize]{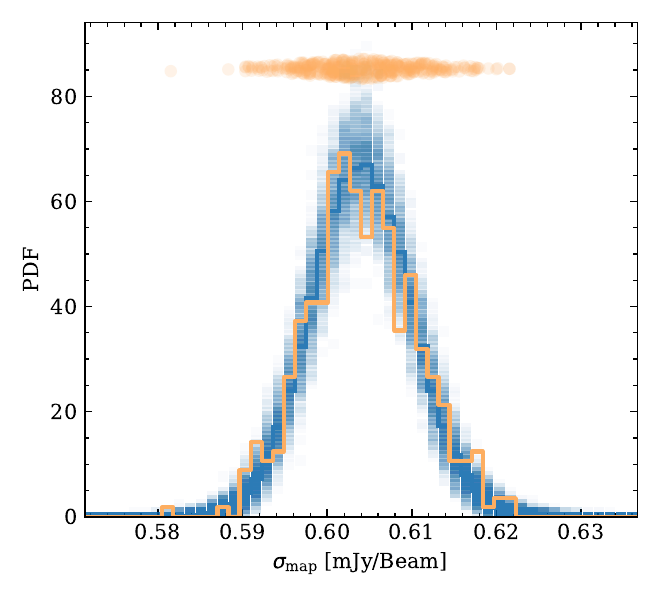}
        \caption{Comparison of the standard deviation $\sigma_{\mathrm{map}}$ measured from different \texttt{simobserve} realizations using 200 different random seeds (orange points and corresponding histogram) and the $\sigma_{\mathrm{map}}$ of the jackknifing for each corresponding mock observation (shaded blue squares). The thick blue line traces the total distribution of the derived standard deviations, $\sigma_{\mathrm{map}}$ when averaging over all the jackknife realizations for all the different \texttt{simobserve} seeds. Through jackknifing, we thus recover the true noise distribution of the observations. }
        \label{fig:test_simobs}
    \end{figure}

    We tested how well jackknifing performs as a function of the sources' $S/N$, as the randomization process might not fully remove the signal, leaving residuals in the generated noise realization. We initially noticed a difference between the standard deviation from the single cleaned map and the mean of the 50 jackknife realizations (see Appendix~\ref{app:sim}). This bias depended on the seeding used in \texttt{simobserve}, which is the result of the inherent variance of \texttt{simobserve}, combined with the fact that the standard deviation is estimated from maps with a finite size. Due to the limited representation of the chosen image domain, large- and intermediate-scale modes can induce non-negligible biases in the measured noise variance compared to the underlying truth distribution. To explicitly test for this, we create jackknifed realizations over a large set of \texttt{simobserve} runs without a source, each assuming a different input random seed (Fig.~\ref{fig:test_simobs}). All the resulting marginalized jackknife distributions are found to be consistent with the same normal distribution, which is characteristic of the noise probability function. The \texttt{simobserve} variance estimates are scattered according to the same distribution, implying that the \texttt{simobserve} output itself shows the same statistical properties of the result of an individual jackknifing cycle. Therefore, jackknifing can be used to better describe the underlying distribution function independent of the noise realization used in \texttt{simobserve}.
    
    \begin{figure}[t!]
        \centering
        \includegraphics[width=\hsize]{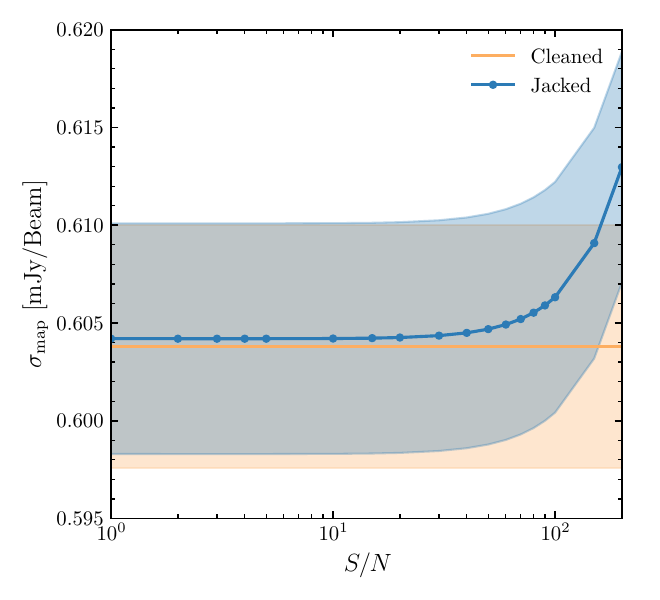}
        \caption{The standard deviation for all maps as a function of the peak $S/N$ of the sources. The blue-shaded region indicates the standard deviation over the various noise estimates from the 50 jackknife realizations for each S/N. The orange-shaded region is the standard deviation over the orange data points in Fig.~\ref{fig:test_simobs}. By jackknifing the observations, we retrieve the noise properties of the cleaned map accurately when the sources have $S/N<50$.}
        \label{fig:sims_vs_snr}
    \end{figure}

    We test how the measured standard deviations from the jackknife realizations evolve with S/N in Figure~\ref{fig:sims_vs_snr}. To this end, we simulated the sources as Gaussians that are off-centered from the phase reference point. This setup allows the signal to leak from the real to the imaginary components, testing the effectiveness of the jackknife routines without assuming the source to be at the phase reference -- an assumption often made in interferometric data analysis tools (such as \texttt{uvcontsub}). We compare the mean and standard error of the standard deviation values measured for the jackknifed realizations vs the observations simulated with \texttt{simobserve}.
    
    We find that our jackknifing approach is perfectly reliable for $S/N \lesssim 50$. That is, the jackknife realizations of simulated observations with sources at $S/N \lesssim 50$ are perfectly consistent with the cleaned noise maps. The observations simulated with a peak $S/N\gtrsim50$, however, indicate a stronger deviation from the cleaned noise estimate. We note that this does not manifest as a residual of the source but as an overall larger standard deviation throughout the map. We therefore advise that before applying \texttt{jackknify} to the measurement set, one should remove extremely bright continuum and line-emitting sources (e.g., $S/N>50$) from the data set in the visibility plane, as is common practice with other line-finding methods \citep[e.g.,][]{Walter2016}. Overall though, the tests we have performed with simulated data (Figures~\ref{fig:test_simobs}~\&~\ref{fig:sims_vs_snr}), show that jackknifing allows for an accurate estimation of the underlying noise distribution.\footnote{We note that we did not evaluate the performance of jackknifing the visibilities in cases of large extended bright emission such as in molecular clouds within low-redshift galaxies. } 
    
    \begin{figure}[t]
        \centering
        \includegraphics[width=0.91\hsize]{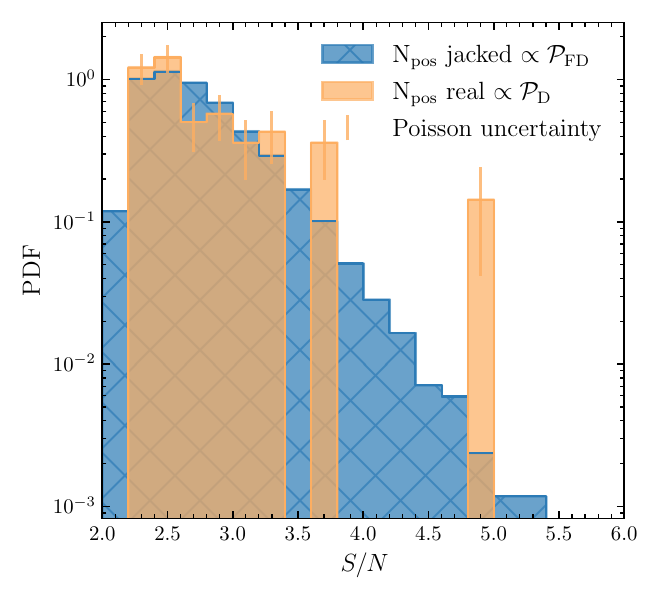}
        \caption{Output of the line finding done on the simulated ALMA observation that had an off-centered line and continuum source which was initialized to have an integrated $S/N$ of 5 in the integrated velocity map (see Fig.~\ref{fig:sims}). In blue, we show the underlying PDF of a false detection $\mathcal{P}_{\rm FD}(x)$ derived from counting peak values in the jackknifed realizations. In orange, we show the results on the simulated data, which shows a clear detection of the two sources at a $S/N=5$. The peak at $S/N\approx3.6$ is consistent with being noise (as explained in the main body of the text)}
        \label{fig:sim_linefinding}
    \end{figure}

    The final validation we performed with the simulated data was to test the accuracy of the line-finding method. Since jackknifed data cubes have an output identical to the original dataset, we apply the same tools to the real and noise realizations. Therefore, we use the simulation for which the sources have an integrated $S/N = 5$. We sample the distribution of false positive detections, $\mathcal{P}_{\rm FD}(x)$, by applying the line finding algorithm to the jackknifed data and use the \texttt{FindClump} algorithm\footnote{We note that this is an arbitrary choice and other algorithms may be better suited} to search for lines in the jackknifed data cubes (see Section~\ref{sec:setup} for a description on how \texttt{FindClump} is setup). The average distribution of the peak values in the 50 jackknife realizations is shown by the blue histogram in Figure~\ref{fig:sim_linefinding}. This histogram reveals a smooth peak distribution, declining from $S/N=2.5$ to $S/N \approx 5.5$.\footnote{We note that the drop-off at lower $S/N$ is due to the cropping of peaks and the way \textit{Source Extractor} catalogs its findings.} Given the number of peaks \texttt{FindClump} detected in a single realization ($\rm N_{\rm peaks} = 70$), and a likelihood of $\mathcal{L}_{\rm FD}(\gamma = 4.5) = 0.0021$, we expect to find on average $0.1 \pm 0.4$ peaks above $S/N=4.5$ in the simulated observation. By applying the line-finding algorithm to the real data (orange histogram), we recover the simulated continuum and lines sources, both with a $S/N = 4.9$. 
    Compared to the underlying noise distribution (in blue), the likelihood ratio tests yield $\Lambda(\gamma = 4.0) = 13$. %
    We also find a small excess at $S/N\approx3.6$ and check if our likelihood ratio test indicates this to be real or not (we know it is not). 
    We find that the number of peaks, expected from jackknifing, at the $S/N$ range of $ 3<\gamma<4$ ($3.5<\gamma<4$) is $14.6 \pm 3.8$ ($2.1 \pm 1.5$), while we find $16 \pm 4$ ($5.0 \pm 2.2$) in the simulated data, resulting in a $\Lambda(3<\gamma<4) = 1.1$ ($\Lambda(3.5<\gamma<4) = 2.4$). Thus, our statistical tests indeed find these peaks to be consistent with being noise, meaning we recover the simulated sources while the noise is correctly characterized as such.
    
\section{Application to $z>10$ galaxy candidates}\label{sec:results}

    \begin{figure*}[t]
        \centering
        \includegraphics[width=\textwidth]{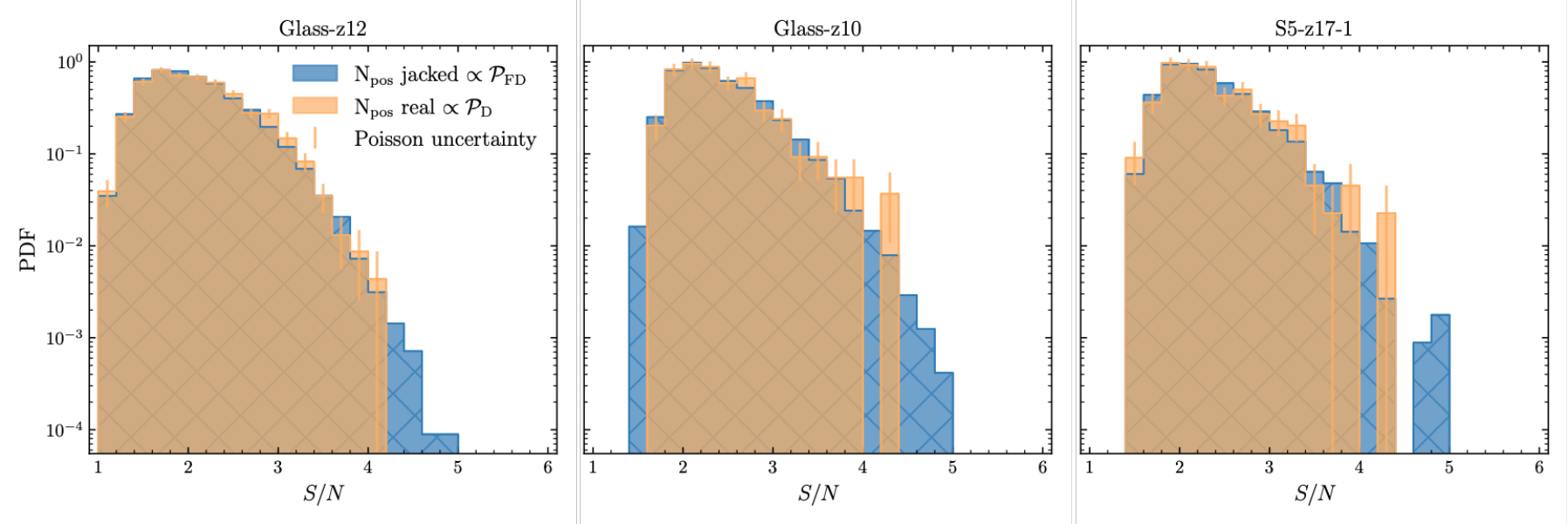}
        \caption{Probability distribution of false detections, $\mathcal{P}_{\rm FD}(x)$ noted as $\rm N_{\rm pos}~$jacked, and the peak value distribution of the observations, $\mathcal{P}_{\rm D}(x)$ noted as $\rm N_{\rm pos}~$real. Both distributions are obtained by running \texttt{FindClump} on the jackknifed cubes and the real data cube, respectively. From left to right, we show the three data sets of GLASS-z12, GLASS-z10, and S5-z17-1 (See Tab.~\ref{tab:observations}). An excess of $\mathcal{P}_{\rm D}(x)$ over $\mathcal{P}_{\rm FD}(x)$ is indicative of line detection. However, no such excess is detected.}
        \label{fig:pdf_compare}
    \end{figure*}
    
    Our tool, \texttt{jackknify}, can be applied to various science cases, including ones involving blind line searches in deep fields (i.e., without known counterparts), targeted searches around candidates identified at other wavelengths (e.g., following up $z>10$ candidates) and searches for companions around known sources. Here, we apply our method to three sets of ALMA observations targeting the $z>10$ galaxy candidates GLASS-z12, GLASS-z10, and S5-z17-1 (\citealt{Bakx2023}, \citealt{Yoon2023}, and \citealt{Fujimoto2023}, respectively). There was also a DDT program targeting the $z\sim 13$ candidate HD1, but we exclude these observations here as more recent work by \citet{Harikane2024b} spectroscopically confirmed the object to be a passive galaxy at $z=4.0$ and a similar jackknifing analysis was already performed in \cite{Kaasinen2023}. Furthermore, we note that during the referee process, two additional ALMA observations targeting $z > 10$ galaxies were reported. Specifically, these include observations on GL-z14 \citep[Project Code: 2023.A.00037.S, PI: S. Schouws;][]{Carniani2024b,Schouws2024} and a deeper observation targeting GLASS-z12 \citep[Project Code: 2023.A.00017.S, PI: J. Zavala;][]{Zavala2024b}. These observations are not included in this work but underscore how rapidly the field is advancing.

    In Section~\ref{sec:setup}, we describe the experimental setups, in Section~\ref{sec:without_prior}, we detail the results of the blind line search across the entire spectral axis and, in Section~\ref{sec:with_prior}, we summarize the line finding results for GLASS-z12, incorporating an additional redshift prior based on spectroscopic JWST/MRS measurements of the H$\alpha$ line \citep{Zavala2024}.
    
    \subsection{Specific experiment set up}\label{sec:setup}       

        To quantify the detection significance of a line along the full spectral axis, we adopt the following procedure.
        \begin{enumerate}
            \item We jackknife each observation set 50 times, creating 50 different noise realizations. Each jackknifing iteration includes differing visibilities and imaging of the full spectral scan and takes $\approx 9$ minutes on a 2019 MacBook Pro with a 2.6 GHz 6-Core Intel Core i7 processor. A tenfold improvement in performance speed was found when using a MacBook Pro with the M3 chip. 
            \item We use \texttt{FindClump} \citep{Walter2016}, as implemented in the \texttt{interferopy} package \citep{interferopy}, to sample the distribution of positive peak values from both the real and jackknifed data.
            \item We mask out features outside a $0\farcs6$ radius of the optically derived center. This corresponds to a physical radius of $\approx 2$~kpc at $z=12$ and is $\approx2\times$ the size of the beam for all four datasets. We chose a radius of $r=0\farcs6$ to include the tentative detection previously reported in our analysis \citep[the feature was found at an offset of $0\farcs51$ from the JWST-derived centroid;][]{Bakx2023}, but be able to exclude serendipitous detections of interlopers. 
            \item We calculate the likelihood ratio of the respective distributions to infer if there is an excess in the peak distribution in the real data compared to the sampled noise distribution.
            \item We consider a detection to be significant if the excess of peaks in the real data with respect to the noise has at least a significance of $2\sigma$ based on Poisson statistics. Therefore, we require $\Lambda(\gamma) k$, with $k=3$. 
        \end{enumerate}

        In step 2, we configure \texttt{FindClump} to search for emission lines with widths between $100 - 500$ km s$^{-1}$, corresponding to $3-11$ times the channel width. The velocity width varies slightly across different observation sets as we are examining galaxies at various redshifts. With \texttt{FindClump}, we identify peaks that exceed a signal-to-noise ratio ($S/N$) of 0, and we crop ``identical'' peaks if they are found within $0.2\arcsec$ of each other spatially (approximately $1/3$ of the beam width) and within 0.2~GHz spectrally (roughly $4-5$ channels apart). Additionally, we have modified the default settings in \texttt{SExtractor} \citep{Bertin1996}, specifically the analysis and detection thresholds which we both set to 2 to better detect low $S/N$ clumps. 
    
    \subsection{Results on blind line searches}\label{sec:without_prior}

        We check for any real signal by comparing the distribution of positive peaks in the real and jackknifed data (Fig.~\ref{fig:pdf_compare}). According to Eq.~\eqref{eq:ratio}, this comparison translates to the ratio of the probability distribution of a detection, $\mathcal{P}_{\rm D}(x)$, versus a false detection, $\mathcal{P}_{\rm FD}(x)$. Thus, any excess of positive features in the real data relative to the noise distribution indicates a true positive line detection. Table~\ref{tab:significances} provides an overview of the likelihood of previously reported tentative detections being real. In the following section, we will discuss the results for each source, one by one.

        \begin{table}[t]
            \centering
            \caption{Significances of Detected Peaks}
            \resizebox{0.95\hsize}{!}{%
            \begin{tabular}{lccc}
                \toprule
                                                                              & GLASS-z12      & GLASS-z10    & S5-z17-1  \\
                Reported significance                                         & $5.8\sigma$   & $4.4\sigma$   & $5.1\sigma$ \\   
                \midrule
                \midrule
                Found highest significance$^\dagger$                          & $4.2\sigma$   & $4.3\sigma$   & $4.3\sigma$ \\     
                $\mathcal{L}_{\rm FD}(\gamma = 4)$                            & 0.0011        & 0.0054       & 0.0032   \\     
                $\mathcal{L}_{\rm FD}(\gamma = 4) \times {\rm N_{\rm peaks}}$ & $1.3 \pm 1.1$ & $1.5 \pm 1.2$ & $ 0.7 \pm 0.8$ \\     
                $\rm N_{\rm peaks}$ found                                     & 1             & 2             & 1 \\     
                $\Lambda(\gamma = 4)$                                         & 0.80          & 1.4           & 1.4 \\     
                \bottomrule
            \end{tabular}
            }
            \tablefoot{In this table, we quote the individual likelihoods of false detections for the three DDT-ALMA data sets on $z>10$ galaxies. Multiplied with the number of peaks in the real dataset (${\rm N_{\rm peaks}}$), this leads to the expected number of peaks above $S/N>4$ in a purely noisy dataset. We compare this with the found number of peaks in the real data and the likelihood ratio test as defined in Eq.~\eqref{eq:ratio}.$^\dagger$ We note that the peaks with the highest significance are not the same peaks as the ones that correspond to the reported significance (top row). See section~\ref{sec:without_prior} for more detail.}
            \label{tab:significances}
        \end{table}

        For GLASS-z12 (left panel, Fig.~\ref{fig:pdf_compare}), we find no excess of positive peaks in the real data compared to the underlying noise distribution. We integrated the probability of a false detection from the detection threshold of $\gamma = 4\sigma$ onward. Considering the total number of peaks in the real data ($\rm N_{\rm peaks} = 1150$), we estimated the expected number of peaks due to noise fluctuations. We found $\mathcal{L}_{\rm FD}(\gamma = 4\sigma) \times {\rm N_{\rm peaks}} = 0.0011 \times 1150 = 1.3 \pm 1.1$. In the real data, we recover one peak above $S/N > 4$, which is consistent with being noise. This is also indicated by the likelihood ratio test (Eq.~\ref{eq:ratio}), which results in $\Lambda(4\sigma) = \mathcal{L}_{\rm D}(4\sigma) / \mathcal{L}_{\rm FD}(4\sigma) = 0.80$. We note that the $S/N = 4.2$ peak discussed here is not the same peak found by \citet{Bakx2023}. The peak we have recovered is at a frequency of $\nu = 246.6106$\,GHz and coordinates RA, Dec $= (3.49895, -30.32469)$, which differs from their reported $5.8\sigma$ peak over 400 km s$^{-1}$, offset $0\farcs5$ from the JWST position (Table~\ref{tab:observations}). We recover a peak at the same location but at a lower significance of $2.9\sigma$ for a linewidth of 280 km s$^{-1}$ (the maximum S/N peak over any linewidth at this position). The difference in S/N and linewidth is due to the difference in how the data were imaged. Whereas we imaged these data using natural weighting and using a channel width of 46 km s$^{-1}$, \cite{Bakx2023} tapered the data to $0\farcs3$ and used a channel width of 150 km s$^{-1}$. This highlights the importance of imaging the jackknifed data in the same way as the real data. 

        For GLASS-z10 (middle panel, Fig.~\ref{fig:pdf_compare}), we find a small excess of peaks in the real versus the jackknifed (noise-only) data at $S/N=4.1-4.3$. Given the noise distribution and the total number of peaks in the real data ($\rm N_{\rm peaks} = 271$), we expect to find $1.5 \pm 1.2$ peaks at $\gamma > 4\sigma$. In the real data, we find two peaks at $\gamma > 4\sigma$. The likelihood ratio resulted in $\Lambda(\gamma = 4\sigma) = 1.4$. This does not exceed our detection threshold of $k = 3$, meaning that the two tentative features\footnote{Both features have a $S/N = 4.3$, a spatial offset of $0\farcs35$ and $0\farcs39$ from the optical counterpart, and are found at $\nu = 291.58$ and $\nu = 296.12$ GHz. Both have a linewidth of 170 km s$^{-1}$.} are consistent with being noise at a likelihood of $\mathcal{L}_{\rm FD}(\gamma= 4\sigma)$ = 0.0054. Neither peak was previously reported by \citet{Yoon2023}. We also recover the previously reported detection at $2.7\sigma$ with a linewidth of 235~km~s$^{-1}$. This line has a false detection likelihood of $\mathcal{L}_{\rm FD}(\gamma=2.7\sigma) = 0.19$ and a likelihood ratio test resulting in $\Lambda(\gamma = 2.7\sigma) = 0.9$.

        For S5-z17-1 (right panel, Fig.~\ref{fig:pdf_compare}),the expected number of peaks is $\mathcal{L}_{\rm FD}(\gamma = 4\sigma) \times {\rm N_{\rm peaks}} = 0.0032 \times 221= 0.7 \pm 0.8$. Given that we find one peak above $S/N>4$ and since $\Lambda(\gamma=4\sigma) = 1.4$, we conclude that these data are also consistent with being noise. We recover the peak reported in \citet{Fujimoto2023}, albeit with a significance of $3.9\sigma$ instead of the reported $5.1\sigma$.  This line has a false detection likelihood of $\mathcal{L}_{\rm FD}(\gamma=3.9\sigma) = 0.0060$ and a likelihood ratio test resulting in $\Lambda(\gamma = 3.9\sigma) = 2.2$.
        
        In conclusion, our analysis of the GLASS-z12, GLASS-z10, and S5-z17-1 datasets reveals no significant detections of any of these (candidate) $z > 10$ galaxies in the ALMA data. All of the peaks identified in the real data are consistent with noise, as indicated by the likelihood ratio tests and comparison with the jackknifed noise distributions. Given the number of peaks per data set, the detection significance does not follow a Gaussian distribution. Since a $4\sigma$ detection should reflect that only one in fifteen thousand random draws should be a false detection, we reach a $5\sigma$ fluke already within a sample size of $\approx 200-1000$ peaks. This highlights the importance of rigorous noise analysis in confirming potential astronomical signals. 
        
    \subsection{Quantifying detection significance with known redshifts}\label{sec:with_prior}
    
        \begin{figure}[t!]
            \centering
            \includegraphics[width = \hsize]{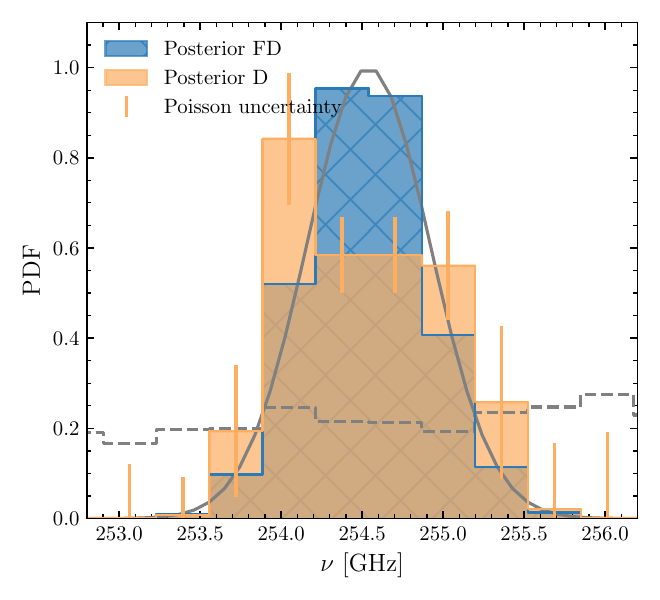}
            \caption{Posterior distribution of peaks above $S/N>2.5$ and within $0\farcs6$ of the JWST derived centroid as a function of frequency for both the real and jackknifed realization. The Gaussian prior derived from JWST/MRS H$\alpha$ observation of \cite{Zavala2024} is shown as the gray line, and the initial PDF of the probability of false detections is the gray-dashed line. }
            \label{fig:pdf_GLASSz12}
        \end{figure}

        So far, we have presented the jackknifing plus line-finding approach for the case of a blind detection experiment and the case where the approximate position of the source is known but where we have no strong prior on the central frequency. We now extend the analysis for cases where the redshift and, hence, the line frequency are known from other observations. To provide an example, \citet{Zavala2024} detected bright nebular emission lines of GLASS-z12 with JWST, constraining the redshift to be $z_{\rm spec} = 12.33 \pm 0.02$. 
        With this prior knowledge, we will reanalyze the data reported in \citet{Bakx2023} 
        at the expected location (spatially and spectrally) of the \oiii line. 
        Notably, this feature is spectrally and spatially offset from the earlier reported tentative detection of \citet{Bakx2023}. 

        The redshift uncertainty ($\Delta z = \pm 0.02$) derived from the JWST/MRS spectrum of GLASS-z12 translates to a frequency width of $\Delta \nu_{\rm obs} = \pm 0.4$~GHz at the expected \oiii frequency of $\nu_{\rm obs} = 254.54$~GHz. We compute the frequency uncertainty from,
        \begin{equation}\label{eq:redshift_to_freq}
            \left|\dfrac{\Delta z}{1+z}\right| = \left| \dfrac{\Delta\nu_{\rm obs}}{\nu_{\rm obs}}\right| = \left|\dfrac{\Delta v}{c}\right|~,
        \end{equation}
        
        \noindent where $\Delta z$ is the redshift range probed, $\Delta \nu_{\rm obs}$ is the corresponding frequency width centered on the frequency $\nu_{\rm obs}$, $\Delta v$ is the corresponding velocity width, and $c$ is the speed of light. This frequency uncertainty is roughly equivalent to the maximum correlated bandwidth of a single ALMA spectral window. Even with better redshift constraints, astrophysical processes like outflows and differing dust geometries could lead to both spatial and spectral offsets, as observed in the \cii and \oiii lines reported by \citet{Fujimoto2024}. Therefore, the likelihood of false detections, $\mathcal{L_{\rm FD}}(\gamma)$, is still determined by the probability density function similar to the blind search scenario, albeit with the addition of a relatively wide prior of $\pm 0.4$\,GHz from the JWST/MRS observations.

        To incorporate the redshift prior from auxiliary data sets into our line-finding routine, we adopt a Bayesian approach using Bayes theorem,
        \begin{equation}\label{eq:bayes}
            P(A|B) = \frac{P(B|A) \cdot P(A)}{P(B)}~.
        \end{equation}
        where $P(B|A)$ is the general expression for a probability function---in our case, the probability of a detection,$\mathcal{P}_{\rm D}(x)$, and false detection, $\mathcal{P}_{\rm FD}(x)$---, $P(A)$ is the prior, and $P(B)$ is the Bayesian evidence. For the detection inference of GLASS-z12, we adopt a Gaussian prior on the redshift, $\mathcal{G}\left(\mu = 12.33, \sigma = 0.02\right)$. We also use a uniform prior on the radial distance, $\mathcal{U}\left(0, 0.6\arcsec\right)$, and a uniform prior on the $S/N$ of the peak, $\mathcal{U}\left(2.5, \infty\right)$, thereby only counting peaks with $S/N > 2.5$. 
        After multiplying the probability distribution by the priors to obtain the posterior distribution, we normalize it such that the integral over the bins of $S/N$, frequency, and radial distance equals one. 

        Unlike methods such as Markov Chain Monte Carlo, which directly sample the posterior distribution, we first sampled the PDFs using \texttt{FindClumps} without applying any priors. We then applied Bayes theorem in post-processing to determine which feature is most likely to correspond to GLASS-z12. This approach assumes no correlation between the $S/N$ of a candidate (since the intrinsic flux of the source is unknown) and its location along the RA, Dec, and frequency axes. This is a reasonable assumption, as there should be no correlation between the $S/N$ of the line detection and these parameters for non-primary-beam-corrected maps.
        
        We visualize our approach to determining the likelihood of a positive peak being real, given the redshift prior, in Figure~\ref{fig:pdf_GLASSz12}. The posterior distributions for both the real and jackknifed observations are shown in the frequency domain. We also show the prior from JWST/MRS observations and the initial PDF of the probability of false detections without applying the prior. The latter indicates that the PDF has a flat distribution with respect to frequency, thus supporting the assumption that there is no correlation between the $S/N$ of a peak and its location along frequency axes. We find a slight excess of detection peaks at a frequency of $\nu = 254.0$\,GHz, corresponding to two peaks located at ${\rm RA}, {\rm Dec}, \nu= 3.49892^\circ, -30.32469^\circ, 253.92$\,GHz and ${\rm RA}, {\rm Dec}, \nu= 3.49918^\circ, -30.32469^\circ, 253.92$\,GHz. These peaks have $S/N$ values of 2.7 and 2.6, and linewidths of 230 km s$^{-1}$ and 140 km s$^{-1}$, respectively. Given the probability of false detection and the number of peaks found in the real data, the likelihood ratio test results in $\Lambda(\gamma=2.5\sigma)= 1.64$. This does not exceed our detection threshold of $k>3$, set in Section~\ref{sec:setup}.

\begin{figure}[t]
            \centering
            \includegraphics[width = \hsize]{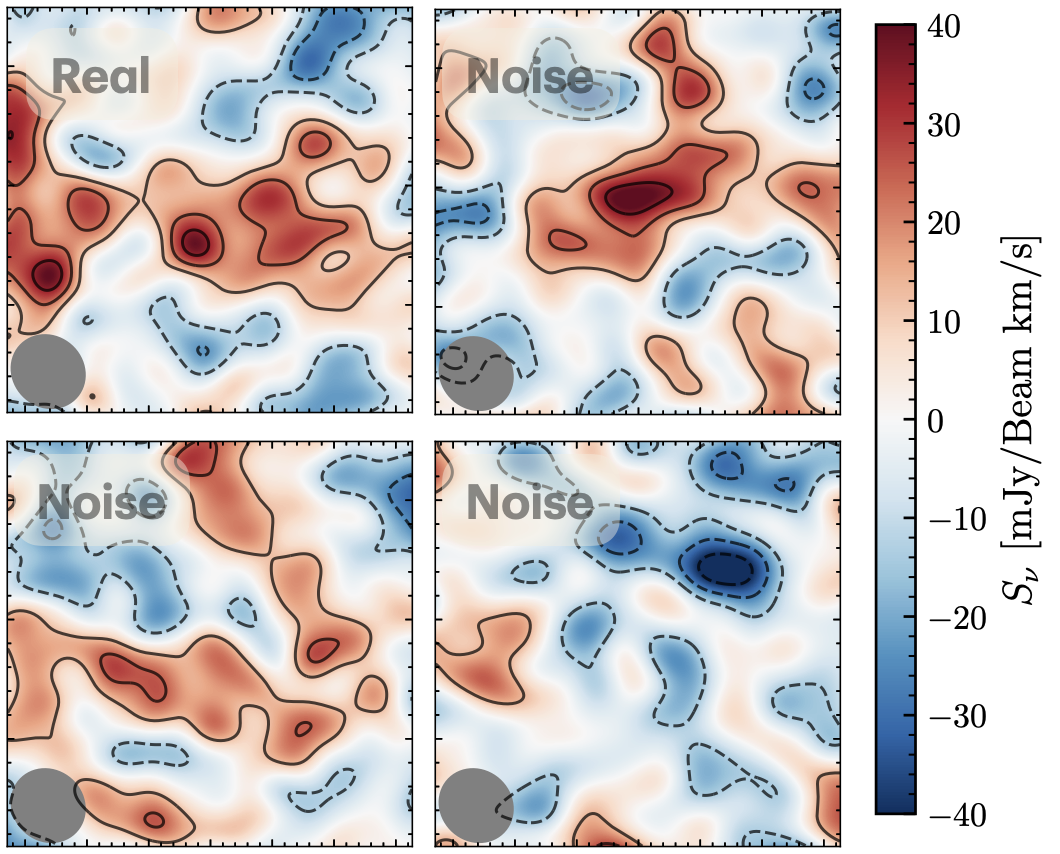}
            \caption{Moment-0 maps centered at $\nu = 254.35$~GHz with a linewidth of 280 km s$^{-1}$ for the real data set and three jackknife realizations which are imaged identically. In the lower left, we show the beam size. The size of the images is $1\farcs7$ by $1\farcs7$. The contours are drawn at $\pm 1,\pm2,\pm3\sigma$.}
            \label{fig:moments}
        \end{figure}

        For illustrative purposes, Figure~\ref{fig:moments} shows a moment-0 map with a tentative feature alongside three jackknife realizations. In one of these realizations, we observe a similar spatial flux distribution as in the real data. The moment-0 maps and the posterior distribution clearly indicate that there is no detection of GLASS-z12 in the ALMA data (Project code: 2021.A0020.S, PI: Bakx). Even including a prior from JWST, the ALMA data is statistically consistent with being noise. To reiterate, we did not include in our analysis the newer observations targeting GLASS-z12 that were recently published  \citep[Project Code: 2023.A.00017.S, PI: J. Zavala;][, see also Table~\ref{tab:observations}]{Zavala2024b}. That work showed a $\approx 4.5-5.2\sigma$ detection of the \oiii line when the new observations were combined with the older data. 
        
\section{Summary \& Implications}\label{sec:summary}

    In this study, we introduce an easy-to-use framework for determining the likelihood of faint emission in interferometric data bring real. By jackknifing the visibilities, we generate noise realizations of the measurement set. Line-finding tools can then be applied to both the noise and the original data set to quantify the level of false detections. In this work, we have tested our approach using \texttt{FindClumps} \citep{Walter2016}, but there are several other line-finding tools, which may be more appropriate for interferometric data, including \texttt{SoFiA} \citep{Serra2014,Westmeier_2021}, \texttt{LineSeeker} \citep{GonzalezL2020} or \texttt{MF3D} \citep{Pavesi_2018}. Our entire analysis procedure, modulo the user's choice of line-finding algorithm, is publicly available at \url{https://joshiwavm.github.io/jackknify/}, and the tool, \texttt{jackknify}, is installable using pip\footnote{\url{https://pypi.org/project/jackknify/}} \citep{jackknify2024}.

    We applied this methodology to three ALMA-DDT observations targeting galaxy candidates at $z>10$ to recover the likelihood of peaks at a certain S/N being real.
    Given the number of peaks in each dataset and their respective distribution of peak values, the likelihood ratio test using a minimum detection threshold of $\gamma=4\sigma$---as formalized in Eq.~\eqref{eq:ratio}--- resulted in $\Lambda(\gamma=4\sigma) = 0.80, 1.4, 1.4$, respectively (see Table~\ref{tab:significances}). Since we set the detection threshold to $\Lambda(\gamma)>k$ with $k=3$ (see Section~\ref{sec:setup}), we conclude that none of the previously reported tentative detections could be distinguished from noise, even when incorporating additional prior information from JWST/MRS slit measurements in the inference.
    
    Our analysis shows that, given the current data volumes ($\rm N_{\rm peaks} \approx 200-1000$), we expect to find approximately $3 \pm 2$ line features with $ S/N \sim 4-5$ in broad ALMA line scans based on the underlying noise distribution. Ensuring a secure detection is therefore challenging, requiring $>5\sigma$ detections when performing a blind line search in a cube probing 30~GHz of bandwidth and targeting a single line. Detecting two lines at matching redshifts would strengthen the significance; however, even then, the likelihood of detecting two noise features within a reasonable spatial and frequency offset needs to be accounted for. Even when probing multiple lines in different bandwidths, spurious features can arise at a significance of $\approx 4\sigma$ at the right frequency and realistic spatial offset in both data sets \citep[e.g.,][]{Kaasinen2023}.

    While the elegance of \texttt{jackknify} lies in its straightforward implementation, new approaches are needed to optimize line searches computationally and to increase fidelity. There is currently no publicly available line-finding algorithm implemented in $uv$-space for extragalactic sources (see, for instance, \citealt{Loomis2018} with the code \texttt{VISIBLE}, which they used for finding isotopes in protoplanetary disks in high spatial and spectral resolution observations). Operations in $uv$-space  are not affected by side lobes or other sources of correlated noise to which analyses in the image plane are susceptible, but they come at the cost of computational efficiency. The flux of the source is not concentrated within a resolution element but rather spread over a large number of visibilities. Additionally, ongoing ALMA studies -- albeit still in the image plane -- are exploring the use of unsupervised machine learning to identify faint emission lines \citep[see, e.g.,][]{Baronchelli2021, Baronchelli2024}. Although these methodologies are in the early stages of development, their implementation could drastically increase computation speed and efficiency.

    Improving current line-finding methods is important in the context of several upcoming surveys and telescope upgrades. For example, the Wideband Sensitivity Upgrade of ALMA (\citealt{WSU2023}) will increase the bandwidth and sensitivity by a factor of 2--4. This upgrade will greatly increase the efficiency of line-finding experiments, especially at high frequencies, and is, therefore, critical for identifying $z>7$ galaxies. 
    Looking to the more distant future, significant advances are expected from major new submillimeter-to-centimeter facilities, such as the proposed 50\,m single dish, named the Atacama Large Aperture Submillimeter Telescope (AtLAST; \citealt{Booth2024}) and SKA \citep{Dewdney2009}. AtLAST would enable large, unbiased surveys of cosmological volumes in multiple bands through the use of on-chip spectrometers \citep[e.g.,][]{Endo2012} and its large $2^\circ$ field of view, providing secure line identifications for large samples of high-$z$ galaxies. Although splitting in the visibility plane will not be possible for single-dish facilities, a similar approach in which the differencing is done in the time domain will still be useful \citep[see, e.g.,][]{Weiss2009}. At centimeter wavelengths, deep HI surveys are already revealing new HI detections up to $z\sim0.4$ (e.g., \citealt{Baker2024}; \citealt{Xi_2024}; \citealt{Kazemi-Moridani2024}), with the full SKA likely to push this to $z\sim 1$. With the increase in data volume associated with these new and upgraded facilities comes the increased potential for spurious line detections. Thus, it is crucial that the community accurately identifies the probability of false detections using statistically motivated approaches like the one we presented in this work.

\begin{acknowledgements}

     We want to thank Dr. Seiji Fujimoto for his efforts as PI in requesting the ALMA data used in this project and for his helpful discussions, which greatly improved the quality of the interpretation of the data herein.
    
     The project leading to this publication has received support from ORP, which is funded by the European Union’s Horizon 2020 research and innovation program under grant agreement No 101004719 [ORP].

    TM acknowledges support from the AtLAST project, which has received funding from the European Union’s Horizon 2020 research and innovation program under grant agreement No 951815.

    This paper makes use of the following ALMA data: ADS/JAO.ALMA\#2021.A.00020.S, ADS/JAO.ALMA\#2021.A.00023.S, and ADS/JAO.ALMA\#2021.A.00031.S. ALMA is a partnership of ESO (representing its member states), NSF (USA) and NINS (Japan), together with NRC (Canada), MOST and ASIAA (Taiwan), and KASI (Republic of Korea), in cooperation with the Republic of Chile. The Joint ALMA Observatory is operated by ESO, AUI/NRAO and NAOJ.

    JvM acknowledges support from R. Willems for making `jackknify` publicly accessible. 
    
\end{acknowledgements}   
\vspace{5mm}

\bibliographystyle{aa}
\bibliography{z10_paper}

\appendix

    \section{Further comparison between the simulated noise and jackknifed realizations}\label{app:sim}
    
\begin{figure}[h]
        \centering
        \includegraphics[width=\hsize]{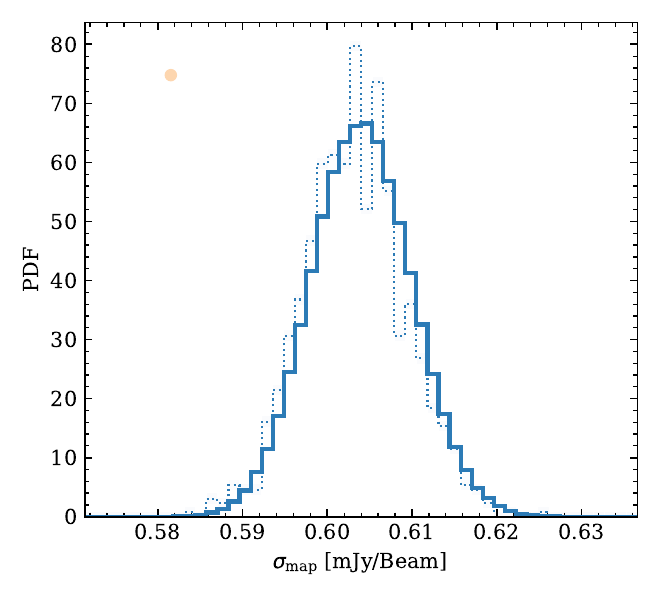}
        \caption{Comparison of the standard deviation $\sigma_{\mathrm{map}}$ measured from a single \texttt{simobserve} simulation (i.e., using one random seed; orange point also shown in Fig~\ref{fig:test_simobs}) and the output of jackknifing that single realization (blue dotted line). The latter clearly follows the total distribution of the derived standard deviations (blue solid line of Fig.~\ref{fig:test_simobs}) instead of being centered on the single \texttt{simobserve} realization.}
        \label{fig:bias_simobserve}
    \end{figure}

    Figure~\ref{fig:bias_simobserve} shows the difference between the standard deviation from the cleaned map of the first, single simulated observation (i.e., one seed used for \texttt{simobserve} visualized with the orange point) and the distribution of standard deviations obtained from the various jackknife realizations (blue dotted line). We find that the jackknifed realizations follow the average distribution of standard deviations, obtained with various \texttt{simobserve} seeds (solid line, Fig.~\ref{fig:test_simobs}), but their median clearly differs from the first single \texttt{simobserve} estimate. This shows that jackknifing can be used to better describe the underlying distribution function than a single noise realization used in \texttt{simobserve}.
        
\end{document}